\documentclass{IEEEtran}
\usepackage[utf8]{inputenc}
\usepackage{amsmath}
\usepackage{graphicx}
\usepackage{amssymb}
\usepackage{array}
\usepackage{wrapfig}
\usepackage{tikz}
\usetikzlibrary{automata}
\usepackage{multirow}
\usepackage{breqn}
\usepackage{dirtytalk}
\usepackage{tabularx}
\usepackage{xcolor}
\usepackage{tabularx}
\usepackage{cite}
\begin{document}
\title{\textbf{Conformal Invariance and Phase Transitions: Implications for Stable Black Hole Horizon?}
}

\author{Pradosh Keshav MV{*} \\
Arun Kenath{*} \\
{*}Department of Physics and Electronics \\
Christ University, Bangalore, India 560029
}
\date{}
\maketitle

\begin{abstract}
The behaviour of black hole horizons under extreme conditions—such as near collapse or phase transitions—remains less understood, particularly in the context of soft hairs and Aretakis instabilities. We show that the breakdown of conformal symmetry during the balding phase induces a topological reorganization of the horizon, leading to divergent entropy corrections and emergent pressure terms. These corrections exhibit universal scaling laws, analogous to quantum phase transitions in condensed matter systems, with extremal limits functioning as quantum critical points. Interestingly, by employing quasi-equilibrium boundary conditions, one could stabilize horizon dynamics without explicitly introducing ad hoc higher-order corrections, further limiting the universal applicability of conformal invariance in black hole physics.
\vspace{3mm}\newline
\textbf{keywords:} Conformal invariance; Phase transitions; Soft hair; Wald entropy; Aretakis instability

\end{abstract}
\section{Introduction}
Since Hawking in \cite{hawking1976} hinted at a potential information loss in black holes through what is now known as \say{Hawking radiation,} they have become a crucial testing ground for reconciling quantum mechanics, general relativity, and thermodynamics. Efforts are ongoing to redefine boundary conditions and resolve the information paradox, as well as to explore the implications of the soft-hair conjecture, as discussed in \cite{arkani2007, goon2020, bleicher2011}. However, the discovery of black hole pressure by Calmet and Kuipers adds another layer of complexity to this paradox \cite{calmet2012}.

The soft-hair conjecture proposes that black hole properties can extend beyond just mass and charge, incorporating additional degrees of freedom (often referred to as \say{hair}). However, decomposing curvature down to the Planck scale and achieving a unique quantum vacuum in this context remains a hypothetical scenario \cite{Faraoni2017}. It follows that understanding the dependence between soft-hair symmetries and black hole horizon stability—especially when incorporating higher-order pressure corrections—could be crucial for resolving the information paradox \cite{Grumiller2020, Haldar2023}.

While investigating the instabilities associated with the translation invariance along extremal horizons, Aretakis \cite{aretakis2012a} demonstrated that the scalar wave equation—when incorporating the decomposed curvature from a generalized wave functional—could exhibit blow-up behaviour due to the need for higher-order derivatives. Aretakis focused on these spacetime instabilities in extremal black hole horizons under scalar perturbations. It is important to distinguish these instabilities from those related to the global symmetries of specific spacetimes, such as the Kerr-Newman and Majumdar-Papapetrou solutions, which were also examined by Aretakis \cite{aretakis2013a}. In this context, the emergence of instabilities linked to translation symmetry presents an interesting area to explore within the soft-hair conjecture framework.

At null infinity, the Bondi-Metzner-Sachs (BMS) symmetry characterizes the asymptotic symmetry of flat spacetime. In \cite{hawking2016}, Hawking bolsters the soft-hair conjecture by demonstrating that a symmetry arising from translation invariance remains BMS invariant. This invariance occurs because the initial data at future or past null infinity undergoes nontrivial modifications—specifically, the addition of extra terms involving second-order derivatives in the effective action. Notably, Hawking’s original argument (1976) suggested that after complete black hole evaporation, the quantum states converge to a unique vacuum, potentially linked to the black hole’s total Wald entropy. However, there remains debate over whether black holes fully evaporate or instead settle into an extremal state at future null infinity \cite{marolf2017, scardigli2020}. In the latter scenario, one must consider the extremal limits—viewed as quantum critical systems—where divergent entropy corrections and pressure terms exhibit universal scaling laws analogous to those observed in quantum phase transitions \cite{Foit2015Self-similar, Cai1997Critical}.

In contrast to Hawking’s later argument, Calmet and Kuiper emphasize understanding horizon stability during the balding phase—when a black hole loses mass and settles into a unique vacuum state. They analyze the dynamics of this phase by introducing the concept of black hole pressure, arguing that quantum gravity corrections to the Wald entropy (a measure of black hole disorder) also yield an effective pressure term. This pressure term can be interpreted as the black hole’s influence on its surroundings. Consequently, under suitable approximations, the horizon data (constructed using higher-order pressure corrections derived through a specific mathematical technique applied to the boundary data) can provide stable boundary conditions, potentially offering advantages over traditional methods that rely on effective field theories or higher-derivative curvature terms, which might not be compatible with a radiation-dominated universe \cite{matsui2019}.

Aretakis instabilities are particularly relevant in this context because they arise from the system’s inherent translation symmetry. By examining these translations, one can potentially delineate regions of horizon instability in a Schwarzschild black hole, especially under extremal conditions where the mass limit is reached. For instance, if the mass change is zero, this may signal a point at which the symmetry is broken. Such a breakdown corresponds to a global symmetry during a phase transition driven by a change in mass and an associated scalar field. Moreover, these transitions are often linked to a non-perturbative breakdown of conformal symmetry  \cite{carlip2007symmetries, almheiri2016conformal}, with the associated effective action yielding the black hole's entropy. This understanding supports the view that conformal symmetry is only broken in the presence of soft hairs, where additional black hole properties beyond mass and charge emerge from the spontaneous breaking of a global symmetry.

In contrast to this view, we investigate the role of conformal symmetry breaking in black hole phase transitions, particularly in the context of horizon stability. We propose that the breakdown of conformal symmetry during black hole evolution leads to a topological reorganization of the horizon, akin to phase transitions observed in condensed matter systems. Specifically, we examine how entropy corrections and emergent pressure terms obey universal scaling laws analogous to quantum criticality in low-temperature systems. This challenges the applicability of a universal conformal symmetry in black hole physics and highlights the need for an alternative mechanism for regulating instabilities associated with extremal horizons.

\section{Quasi-Equilibrium Boundary Conditions for Black Hole Horizons}
The conformal thinsandwich decomposition serves as a robust approximation for constructing quasi-equilibrium data. However, the described phase transitions rely on initial Euclidean data, which is highly asymptotic. The theorem developed by Pfeiffer and York in 2005 surprised many with its original form due to its non-uniqueness in solving independent background data. The theorem was discovered to contain two branches, one with relatively low gravitational effects and the other leading to singularity \cite{baumgarte2007}. York popularized and widely used this variation of the original theorem (non-conformal) for modelling binary objects in astronomy.

Furthermore, considering the variability and constraints imposed by the conformal thinsandwich decomposition, \cite{cook2008} pointed out that quasi-equilibrium boundary conditions were much more useful for modelling quasi-data. Their results were supported \cite{becker2003} by the asymptotic expansion of the bulk just outside the black hole horizon. This region seems to coincide with the initial data of a space-like hypersurface, equivalent to the boundary data of the outermost trapped surface of the black hole's apparent horizon. 

In this particular setup, three approximations are proposed indeed to track the evolution of the quasi-boundary data, which is otherwise challenging to theorize without quantum gravity corrections:
\subsection{Vanishing expansion and shear stress}
The first approximation implies an expansion of scalar $R$ with the space-time metric $g_{\alpha \beta}$ vanishing at a three-surface $\Sigma^{(3)}$:
\begin{equation}
    R \mid_{\Sigma^{(3)}} = 0.
\end{equation} In dynamical regimes such as the balding phase, this condition is justified by the suppression of expansion-driven instabilities under quasi-equilibrium (see \cite{york1983dynamical} for staticity assumptions in York’s formalism). For any isolated system, we assume that the unstable region is space-like $(s_i =s_1 + s_2 \dots )$, satisfying the boundary conditions for scalar data $\Sigma_S$ as pointed out by \cite{goon2020}. This data effectively replicates regions with the initial conditions similar to the apparent horizon of a black hole. The outgoing orthogonal null geodesic $k^\mu$ also vanishes at the boundary where the viscous shear stress $\sigma_{\mu \nu}$ vanishes at,
\begin{equation}
    \sigma_{\mu \nu}\mid_{\Sigma^{(3)}}  = 0
\end{equation}i.e., the trapped surface where the shear is orthogonal to a three-surface of quasi-equilibrium boundary data. The vanishing shear is indicative of the loss of rotational distortion as angular momentum \(J \to 0\), aligning with the axisymmetric-to-spherical topological transition. The particles that enter the region would be causally disconnected with a linear mapping:
\begin{equation}
   \Sigma_{i}^{(3)} s_i  \xrightarrow{\tau} \Sigma_S = \frac{\boldsymbol{G}}{4 \pi^2  H(r)},
\end{equation} 
where the torque $\boldsymbol{G} \to 0$ vanishes at the boundary, or else, $H\geq 0$ is the horizon's half thickness because of the loss of angular momentum during the balding phase. Conventionally, \(\boldsymbol{G} \propto J\) represents the angular momentum flux across the horizon boundary and \(H(r) \sim r_s\) (Schwarzschild radius). As with the (quantum) scaling effects that act on surface $S$ in equation (3), we assume that $\boldsymbol{G} = J/\tau$ where \(\tau\) is the timescale of the balding phase. This corresponds to a change where the horizon transitions from an axisymmetric (\(S^1 \times S^2\)) to a spherical (\(S^2\)) topology using a true evolution parameter ($\tau$) acting on a non-universal function $\lambda(v)$ (which is a dimensionless conformal scaling factor) as well as on the azimuthal components of $s_i$.

For a particle freely falling from infinity in the Schwarzschild metric, \cite{lemos1995} showed the exact solutions of Einstein's field equations with consistent cylindrical symmetry
\begin{dmath}
    ds^2 = - \left[\left(\lambda^2 - \frac{\omega^2}{a^2}\right)a^2 r^2 - \frac{b \lambda^2}{ar}\right]dt^2  - \frac{\omega b}{a^3 r}d\phi dt +
  \frac{dr^2}{a^2 r^2 - \frac{b}{ar}}+ \left[\left(\lambda^2 - \frac{\omega^2}{a^2}\right) r^2 + \frac{\omega^2 b }{a^5 r}\right]d \phi^2 +a^2 r^2 dx^2,  \label{eq4}
\end{dmath}
for the limits \( -\infty < t < \infty, \  0 \leq r < \infty, \ 0 \leq \phi <2\pi, \ \text{and} -\infty < x < \infty\). Here, the term $ds^2$ represents the infinitesimal distance between two neighbouring points in spacetime, with free parameter $\lambda$ included in the time-time and azimuthal-azimuthal component of the metric tensor along the spatial coordinate in the $x$-direction. Also, \(a = -\Lambda/2\) invokes cosmological constant effects, with \(\omega\) parametrizes angular momentum, and \(b\) arises from integration constants in Einstein’s equations. 

The cylindrical symmetry in equation (\ref{eq4}) can have implications for equation (3) when $\Sigma_S$ acts as a Riemannian covering of $\Sigma^{(3)}$ within the limits of the radial circumferential coordinate $0\leq r<\infty$. Suppose, if we take  $a= - \frac{1}{3}\Lambda < 0$ for an expanding fluid with free parameter $\Lambda$, such that the particle is at inertial rest from a distance during the coordinate transformation:
\begin{dmath}
  \frac{d\Vec{u}}{d \tau} \to \frac{d\Vec{v}}{d \tau} {:=} \lambda(u) \frac{d}{dt} \lambda(u) (c, u_x, u_y , u_z) \to \lambda(v) \frac{d}{dt} \lambda(v) (c, v_x, v_y , v_z).
\end{dmath} This means that an appropriate coordinate transformation for $R$ creates inertial coordinates for the particle to stay in the same spatial positions during phase transitions. Moreover, similar to equation (3), the null data of the particle also vanishes at $\Sigma_{;k}$ during the transformation—the semi-colon represents the second-order derivatives with unit normal $\mu$ to the surface. This is a useful a priori to tackle the second-order quantum corrections of metric $g_{\alpha \beta}$ when the thickness of the membrane $H(r)$ also vanishes at the polynomial singularity $ar=0$ as \(\lambda \to 1\).
\subsection{Symmetry Breaking in Apparent Horizon}
The second approximation considers the trapped region to be a second-order real space $\mathcal{R}_{(2)}^4$, which is prompted by the initial data:
\begin{equation}
    {\mathcal{L}}_{;k} H(r) \mid_{\mathcal{R}} = 0,
\end{equation} 
where ${\mathcal{L}}$ is the Lie derivative for the null congruence described by Raychaudhuri's equation for the geometrical flow of particles, and the surface curvature $\Dot{S}$ evolves with time $t$ as follows:
\begin{equation}
   \mathcal{L}_{;k} \Dot{S}(t,x,y,z) = -\frac{1}{2} \Dot{S}_{\alpha \beta}\Dot{S}^{\alpha \beta} - \sigma_{\mu \nu} \sigma^{\alpha \beta} + dM H^{\mu}H_{\mu}.
\end{equation} 
Else, $H^{\mu}$ (\(H^{\mu}=\nabla^\mu M\), representing mass flux) has non-linear relations with the conformal four-by-four transformation matrix $\Dot{S}_{\alpha \beta}=\Phi^* \Hat{\Dot{S}}_{\alpha \beta}$, which is assumed to be at inertial coordinates invoking unstable regions in the surface. This is because there are no free parameters apart from the mass $M$ in the Schwarzschild black hole. Even if we change $H(r)$ and conformally transform the surface data $s_i$, the first-order phase transitions remain the same. 

Hence, for simple purposes, we can assume a global symmetry out of the unstable space-like regions in the form
\begin{equation}
    \Dot{S}_{\alpha \beta}\Dot{S}^{\alpha \beta} = \sigma(x,y)\left(\frac{(\Hat{s}_1 + \Hat{s}_2 + \dots)(N^2)}{dM}\right) \Hat{\Dot{S}}_{(-)},
\end{equation} 
such that the shear $\sigma(x,y)$ vanishes with symmetry breaking $(\Hat{\Dot{S}}>0)$ at the apparent horizon of a Schwarzschild black hole and enters into a coordinate transformation of the space-time metric $g_{\alpha \beta}$, which is at relative rest. This transition is topological as it involves discontinuous changes in global symmetry: the axisymmetric structure of the spinning black hole (in the Killing vector $\partial_\phi$) evolves to spherical symmetry, definitively altering the horizon’s Euler characteristic. Soft hairs, represented by BMS charges at null infinity, act as topological invariants that distinguish these phases. However, these charges vanish at the post-balding phase (\(J \to 0\)) while transitioning to spherical symmetry.

Note that the degrees of freedom induced by the four-matrix will also change at the boundary. For a spherically symmetric Schwarzschild black hole, analogous to the four-matrix, we only take the twist of congruence as zero with the smooth covering of $\mathcal{R}^{4}$. 

In the surface, $\Dot{S}$ comprised of bulk represents asymptotic regions, a null geodesic $k^\mu$ satisfies $\Sigma_\mu k^\mu = 0$. This scenario indicates that the trapped region eventually loses mass through Hawking and Unruh-Starobinski radiation. However, this is not the case of generality if we consider a bulk composition of space-like regions
\begin{equation}
    \Sigma_{;k} R_{\mu \nu}\sigma(x,y) \leq dM + 2\lambda \frac{16 \pi}{G M^2} dM,
\end{equation} 
where $G$ is the gravitational constant, $dM$ is the mass of asymptotic bulk, and $\lambda$ is a quasi-static component coming from quantum gravitational corrections. \par

As we can see, in equations (8) and (9), $dM$ vanishes due to the change in gravitational entropy per unit mass, and the translation symmetry $u_x \xrightarrow{}v_x$ varies accordingly. By fixing the three killing vectors $\frac{\partial}{\partial x}$—translation symmetry along the axis, $\frac{\partial}{\partial \phi}$—closed periodic trajectories around the axis, $\frac{\partial}{\partial t}$—invariance under time translations, one can assume that the vanishing space-time metric follows equation (4) at $\Delta\Dot{S}=0$, similar to a "super Penrose process" \cite{liu2020}. This process commonly leads to a reduction of gauge hair pair production in the azimuthal component of the metric, eventually becoming extremal at $\Delta_g = 0$, i.e., $r=M \pm \sqrt{M^2 - \omega^2}$ and $\lambda$ becomes independent from the time component of $g_{\alpha \beta}$.

The two killing vectors $\frac{\partial}{\partial \phi}$ and $\frac{\partial}{\partial t}$ corresponding to the Cauchy's horizon and the event horizon of the black hole also become extremal when both horizons coincide. The evolution of horizon membrane can be determined by tracking the time vector $t^\mu$, i.e., the null congruence of $\Sigma_{;k}$ in parameterized space-like regions, which will be discussed further in section 3. The membrane's mean curvature flow of hyper-surface interiors is as peculiar as the general BSW mechanism for a freely falling particle with no spin. 
\subsection{Lyapunov Functional and Deformed Cauchy Ensemble}
The third approximation takes the Lyapunov functional of hyper-surface stability \cite{zhu2016}, which maps the curvature flow of the membrane to a quasi-equilibrium boundary. This could be done by projecting the surface of $\Dot{S}$ to its space-like regions via a deformed Cauchy's random ensemble and large phase transition to form a marginally trapped region dependent on the thickness of the horizon membrane and independent elsewhere. 

For the potential $H(x)=H(y)=0$, as in the Penner model \cite{Russo2020}, the deformed Cauchy's ensemble is given by:
\begin{equation}
    H(x) = A \ln(x^2 +1) + \lambda (x^2 +1 + \sigma) - \lambda (x^2 +1).
\end{equation} 
When $H(x), H(y) > 0$, equation (9) is called the deformed Cauchy's ensemble, and $A=N$ is our area of interest, which describes the unitary evolution of the ensemble with lapse $N$. One could also observe that \(\lambda\) acts as a control parameter in equation (10) analogous to temperature in Landau-Ginzburg theory, driving phase transitions via horizon energy landscape modifications. 

To put out the latter scenario precisely, the ensemble neither contracts nor expands when the apparent horizon becomes extreme. The membrane $H(r)$ evolves under the boundary data $\Sigma_{\mu}K^\mu$. The surface of the membrane could be assumed to be a replica of the mean curvature flow of an accelerator. 

At the boundary, the expansion is stabilized, and the convergence of an extrinsic curvature $K$ is given by:
\begin{equation}
    A_{+}= - \text{tr}(K) + K(\mu , \mu) - (N-1) H
\end{equation} 
and, 
\begin{equation}
  \hat{A}_{-}= \hat{A}_{+} +(2N-1)\hat{H},  
\end{equation}
where $\mu$ becomes the exterior unit normal of $dM$, and $dM$ becomes the boundary if $\hat{A}_{+}=0$. 

\section{Semi-classical Phase Transitions}

The null geodesic congruence of the soft hair conjecture is anticipated to vanish due to the topological defects of black holes. Despite Aretakis' arguments, the black hole is expected to achieve finite-scale stability at $H=0$, termed the Planck relics \cite{barrau2019, macgibbon1987}. However, according to Calmers and Kuipers, an observer at an inertial rest may track back the geodesic if the horizon data contains terms representing Wald entropy.

In \cite{maxwell2005}, it is demonstrated that an asymptotic Euclidean data $(M,g, K)$ satisfying the conditions:
\begin{equation}
\begin{split}
     R - \mid K \mid^2 - \text{tr}(K^2) &= 0, \\
    - \text{tr}(K) + K(\mu , \mu) - (N -1)H &= 0, 
\end{split}
\end{equation}
on $dM$ serves as functional data where the boundary is not a marginally trapped surface when $R$ is a constant scalar. It indicates that $dM$ represents a trapped surface only if $\hat{H} \leq 0$, given a mapping from the future null infinity of an unstable boundary to past null infinity. This choice of mapping is based on the assumption that neither an ongoing particle nor an outgoing particle will leave the surface without a trace in the black hole's entropy.

By deriving $H(r)$ assuming the metric $ds^2$ is retarded in coordinates $(u, r, \theta, \phi)$ near the future null infinity $\hat{A}_+$, which reads $ds^2 = -dt^2 + (d x^{i})^2$, we can relate the standard Cartesian coordinates by $r^2 = x^i x_i$ and $u=t-r$ where $u$ is the retarded time. In advanced coordinates $(v, r, \theta, \phi)$ near past null infinity $\hat{A}_-$, we have $v=t+r$ and $x^i = \hat{x}^i (\theta , \phi)$. Thus, in functional terms, $\hat{H} \geq 0$, if $\sigma(\mu, \mu) \leq 0$, implies the data:
\begin{equation}
 (N-1)\hat{H} = \hat{K}(\hat{\mu}, \hat{\mu}) = \Phi^{*} \ \sigma(\mu, \mu),
\end{equation}
where $\Phi$ is a conformal factor tending to 1 at infinity. Under the boundary data $\sigma(\mu , \mu) \geq 0$, it is also possible to construct an apparent horizon with a marginally trapped surface if $\sigma(\mu , \mu) = 0$, which then shows the background $\hat{A}_{-}=0$ and $\hat{H}=0$. This distance $ds^2$ can be made finite in eq(4) using a four-matrix $S_{\alpha \beta}$ such that $x^{\mu} \xrightarrow{} x^{\mu^{'}} = x^{\mu} + H^{\mu}$, where $H^{\mu}$ is a set of four mixed numbers, and the translations leave $\Delta x$ unchanged over $t$. Thus, by constructing a metric translation using equations (3), (8), and (10), we can track stable boundaries for phase transitions from $g$ to $\hat{g}$ satisfying $R=0$ for all $H \leq 0$. 

Further, by using the extremal conditions of the metric, we could also limit the Cauchy's ensemble as 
\begin{equation}
    \hat{A_{ij}} = \frac{1}{2\hat{N}}(\hat{L}_{;k}\dot{S}(x,y)-\hat{\mu}_{ij}),
\end{equation} 
where the lapse $N=\Phi^{*}\hat{N}$ is bounded by conformal lapse $\hat{N}$. By inverting the coordinate transformations in equation (14), we also get the static space-time, which is assumed to be asymmetric because of quantum fluctuations, unstable at the extremal boundary $N=A_{+}(A_{-})$.

\subsection{Phase Transition: Balding Phase}

This phase is assumed to stabilize hypersurface data by mapping null infinity and quantum corrections to $s_i$. The phase also elucidates the asymmetric nature of unstable space-like regions and their rectification using initial data [$R=0; \ H \leq 0 ; \ \sigma(\mu , \mu) \leq 0$] derived from black hole dynamics, leveraging a Schwinger pair production mechanism.

The action describes an effective Lagrangian dependent on the induced metric $R_{\alpha \beta}$ and fluctuating extrinsic curvature ($K$), denoted by the trace value $K = H^{ij} K_{ij} = K_{i}^{i}$. The intrinsic properties of the surface are fully determined by $H^{ij}$, while $K_{ij}$ dictates how the surface is embedded in a (3+1)-dimensional space. A similar approach by \cite{buonanno1995} underscores that the intrinsic properties of $\Sigma^{(3)}$ follow general covariance and reparametrization invariance within the horizon's volume.

The most general action compatible with the singularity involves considering a planar membrane located at $x=0$, formulated as:
\begin{dmath}
    S_{tt} = - \mathbf{T} \int d^3 \Phi \sqrt{-H} (1+ \Sigma_{;k} {\it P}_{k}^{(1)} {\it Q}_{k}^{(1)} + \Sigma_{;k} {\it P}_{k}^{(2)} {\it Q}_{k}^{(2)} \dots ).
\end{dmath}
Here, $\mathbf{T}$ represents the membrane tension, $\Phi$ denotes the conformal scalar, and ${\it P}$, ${\it Q}$ are parameters and operators respectively, with dimensions $dM$, $(dM)^{2}$, etc.

The Nambu-Goto action for strings and the system's action $S_{\text{eff}}= \int L dt$ with Lagrangian $L$ help determine the space-like interval ($\Delta_g$) for the transition:
\begin{equation}
    \Sigma_{;k}\Sigma_{i}^{(3)}s_i: A_+ \xrightarrow{} \Delta_g,
\end{equation} where the radial components of the space-time metric can be deduced from this transition, and conformal symmetry is broken for the balding phase, as discussed in section 2.2.

Deducing $H^{ij}$ involves developing time-independent coordinates for the collapsed particle:
\begin{equation}
    ds^2 = \Sigma_{;k} \int d\Sigma \frac{\partial L}{\partial R_{\mu \nu \alpha \beta}}\mid_{r\geq r_s} - \lambda dt^2 + \Lambda_g,
\end{equation} where $\lambda$ and $\Lambda_g$ represent parameters from time intervals and additional (quantum-gravitational) contributions to the metric tensor, possibly arising from quantum mechanical perturbations, respectively. Here, the first order designates Dirac's membrane action, and the second order designates Nambu-Goto's action.

By combining various projections of $K_{ij}$, which is usually normal to the extremal boundary, quantum fuzzy interiors deformed towards the fluctuating extrinsic curvature with an extremal boundary can be anticipated. The cylindrical constraint in equation (18) is detailed through a local embedding with second-order projections over a smooth Riemannian manifold $\mathcal{R}^4$. Parameters representing the constrained membrane are:
\begin{equation}
    \begin{split}
 Q_{1}^{(2)}=R , \ \ \  Q_{2}^{(2)}= K^2, \ \ \  Q_{3}^{(2)} = K_{ij}K^{ij}.
    \end{split}
\end{equation}
Additional degrees of freedom due to soft hairs allow for tangential deformation of the membrane. For an unconstrained membrane, third-order projections from equation (16) can be formed where each parameter represents the intrinsic constraints of the space-like membrane undergoing deformation:
\begin{equation}
    \begin{split}
 Q_{1}^{(3)}=R_{ij}K^{ij} , \ \ \  Q_{2}^{(3)}= K^3, \ \ \  Q_{3}^{(3)} = K_{ij}K^{ij}K,
    \end{split}
\end{equation} where $Q^{(2)}_i$ corresponds to second-order projections, while $Q^{(3)}_i$ corresponds to third-order projections. Here, the curvature radius $r_s$ embedding in $\Sigma^{(3)}$ characterizes $p$-brane constraints in the Schwarzschild metric. The true evolution parameter $\Hat{\tau}$ satisfying a generalized Schrodinger equation, the intrinsic curvature of $\Sigma^{(3)}$ embedded in a four-dimensional manifold defines lapse $N$ and conformal lapse $\Hat{N}$:
\begin{equation}
    \begin{split}
        \Sigma_{\mu \nu} = \frac{1}{2N} \lambda_{\mu}^{\alpha} \lambda_{\nu}^{\beta} - {\mathcal{L}}_{;\tau} \ g_{\alpha \beta}(Q_{1} +Q_{2}+Q_{3} + \dots), \\
        \Hat{\Sigma}_{\mu \nu} = \frac{1}{2\Hat{N}} \lambda_{\mu}^{\alpha} \lambda_{\nu}^{\beta} - { { \Hat{ \mathcal{L}}}}_{;\Hat{\tau}} \ \hat{g}_{\alpha \beta}  (\Hat{Q}_1 + \Hat{Q}_2 + \Hat{Q}_3 + \dots).
    \end{split}
\end{equation}  Here, we have represented \(\lambda\) as a tensorial conformal scaling factor that distributes the transformation across spacetime directions. However, during the balding phase transition, the shear terms induce deformation and stretching of the membrane as it interacts with the surrounding spacetime. Whereas $\sigma_{\alpha \beta}$ and $\Hat{\sigma}_{\alpha \beta}$ capture the changes in the membrane's shape and geometry due to gravitational effects. The shear of the membrane with ongoing orthogonal null rays is defined by:
\begin{equation}
    \begin{split}
        \sigma_{\alpha \beta} = \Sigma_{;\tau} - \frac{1}{2}\Lambda_{\alpha \beta}N_{-} (P_1 + P_2 + P_3 + \dots),\\
       \Hat{ \sigma}_{\alpha \beta} = \Hat{\Sigma}_{;\Hat{\tau}} - \frac{1}{2}\Lambda_{\alpha \beta}\Hat{N}_{+}(\Hat{P}_1 + \Hat{P}_2 + \Hat{P}_3 + \dots).
    \end{split}
\end{equation}
These equations replicate the boundary data of the Schwarzschild metric, employing a congruence mapping $\Sigma_{;k}$ of conformal decomposition of initial data $\sigma_{\alpha \beta}$ and $\Sigma_{\mu \nu}$ embedded in a three surface $\Sigma^3$ to the vanishing metric $g_{\alpha \beta}$.

\subsection{From Spinning to Balding}
The black hole mapped in the balding phase is not perfectly spherical. The deflection from flat space-time by a geodesic radial null-one-form is achieved at the balding phase, where a spinning black hole loses all its spin and enters a non-spinning state (which wouldn't be a true Schwarzschild black hole due to its history of having spin). The effective action governs this deflection
\begin{dmath}
    S_{\text{eff}} = \int \sqrt{g} \, d^4 \left( \frac{R}{16 \pi G} + P_1(\mu) R^2 + P_2(\mu) R_{\mu \nu}R^{\mu \nu} + P_3(\mu) R_{\mu \nu \alpha  \beta}R^{\mu \nu \alpha \beta} + L \right),
\end{dmath} where $\mu$ represents the unit normal to the surface of $\Sigma^{(3)}$. Here, $P_i(\mu)$ (\(P_1(\mu)=\alpha \mu^2\) where \(\alpha\) is fixed by matching semiclassical entropy corrections) represents the potential functions governing the interaction between the black hole and its surroundings, while $L$ denotes additional Lagrangian terms describing the phase dynamics. Higher-curvature terms such as (\(R^2, R_{\mu\nu}^2\)) in \(S_{\text{eff}}\) are sensitive to global spacetime topology. For instance, the Gauss-Bonnet term \(\int R_{\mu\nu\alpha\beta}^2 \sqrt{g} d^4x\) contributes to the Euler characteristic, tying the balding phase to topological invariants.

Hawking and Gibbson suggested that the laws of thermodynamics could be applied to study black holes using a partition function $Z$ and a path integral:
\begin{equation}
    Z = \int \mathcal{D}(\Sigma^3) \, \exp\left( -\frac{1}{\hbar} S_{\text{eff}} \right),
\end{equation} where $\mathcal{D}(\Sigma^3)$ denotes a functional integration over all possible configurations of the three-dimensional surface $\Sigma^3$, capturing the variability in the geometry and topology of the spacetime manifold. Note that functional integration over horizon configurations \(\mathcal{D}(\Sigma^3)\) mirrors the path-integral formalism for quantum critical systems, where fluctuations dominate at zero temperature. The extremal limit corresponds to a critical point where conformal symmetry breaking in equation (8) invokes scale-invariant fluctuations in the effective action.

It was shown in \cite{elizalde2000} that the effective action of a deformed metric could be defined in Euclidean coordinates:
\begin{dmath}
    ds^2 = -(1-H)dt^2 + 2H \, dt \, dr +(1+H)dr^2 + r^2(d\theta^2 + \sin^2 \theta \ d\phi^2),
\end{dmath}where $H = H^{\mu}H_{\mu}$ is a functional of the radial circumferential coordinate $r$. Curved space-time serves as a useful analogy for this scenario, with a decomposition of $K$ over two killing vectors, namely the time-like $\frac{\partial}{\partial t^\mu}$ and the space-like $\frac{\partial}{\partial \Sigma}$, associated with cylindrical constraints.

The generators at the cylindrical wall are given in terms of second-order derivatives:
\begin{equation}
    K \simeq \frac{\partial}{\partial t^\mu} + 2H \frac{\partial}{\partial \Sigma},
\end{equation} which are usually aimed to align with the first-order projection of $K = H_{ij}K^{ij}$. Assuming that the quantum-gravitational correction term \(\Lambda_g\) is absorbed into the congruence mapping \(\Sigma_{;k}\) under the quasi-equilibrium boundary conditions and by comparing equations (25) and (26) with equation (18), we obtain time-independent coordinates for a collapsed particle:
\begin{equation}
    ds^2 = \Sigma_{;k} \int d\Sigma \, \frac{\partial L}{\partial R_{\mu \nu \alpha \beta}}\mid_{r=r_s} - \frac{\partial}{\partial \Sigma}\lambda \, dt^\mu,
\end{equation}which gives the area scale characterizing p-brane constraints:
\begin{equation}
    \begin{split}
        \frac{1}{Z} K_{ij} &= Q^{(2)}\left(\frac{1}{r}\right), \\
        \frac{1}{Z} H^{ij} &= Q^{(2)}\left(\frac{1}{r^2}\right).
    \end{split}
\end{equation} Comparing the length scale and area scale of p-brane constraints in both cases of equation (28), one also obtains the boundary data [$K, H, R$] analogous to the balding phase that contains the component of Wald entropy and becomes heuristic for all higher-order $S_{\text{eff}}$. The extremal boundary data not constrained with the brane data is an azimuthal component $\Sigma;_{\tau}$ and expansion parameter $\Lambda \neq 0$, which can be traced to its interior components of hyper-surface using equation (28). The unconstrained brane indicates the natural tendency of curvature to deform towards the Euclidean data if $d \Sigma$ is a Gaussian.

Considering that the spin-down phase must preserve traces of rotational history in $\dot{S}$, the membrane can induce an information loss in terms of Hawking radiation and the unit normal $\mu$ to the surface of $D(\Sigma^3)$. This could be explained by the conformal decomposition of the thin membrane $H_{ij}$ and the twisted $\Sigma_{;\tau}$:
\begin{equation}
    \begin{split}
        H_{ij} &= \Phi^4 \hat{H}_{ij}, \\
        \Sigma_{;\tau} &= \Phi^{-2} \hat{\Sigma_{;\tau}},
    \end{split}
\end{equation}where the shear of gaining null rays is given by the quasi-static pseudo-Newtonian form:
\begin{equation}
    \sigma_{\mu \nu} = \Sigma_{;k} - \frac{1}{2}K_{\mu \nu} \lambda_{\alpha \beta} - \Phi^2(r),
\end{equation} where $\Phi(r) = \frac{-GM}{r-r_s}$ represents the gravitational potential per unit mass. If $r_s = \frac{2GM}{c^2}$, then $r_s$ represents the boundary of surface $d\Sigma$ and an excision boundary $r$ in the form of a Robin-type boundary. The area and length scale characterizing p-brane are given by a partition function $Z:d^3 \Sigma \xrightarrow{} \Sigma_{;k}$, which tends to zero with an infinite boundary. This could be explained using the membrane $H(r)$, which is a function of the radial coordinate $r$ and has an asymptotic bulk translation throughout the trapped surface of $\Sigma^{(3)}$.
\section{A hint of Continuum near collapsing objects}
In black hole thermodynamics, extremal limits are often associated with quantum critical points (see Sachdev \cite{sachdev2009quantum} for a holographic description of quantum critical points, with extremal black holes in anti-de Sitter space ) where Hawking radiation and pressure corrections dominate \cite{ali2024quantum}. However, the general scaling approach to collapsing objects indicates a critical point above which a black hole is inevitable. The mass-energy contents for collapsing objects, particularly for spherically symmetric bodies, follow the form $p + \rho = 0$.

At the end of the balding phase, for a Schwarzschild black hole, the entropy and action are related as per \cite{badiali2005}:
\begin{equation}
    \frac{d S_{\text{eff}}}{k_B} = \ln dZ,
\end{equation}implying that the evolution of $S_{\text{eff}}$ determines the boundary change for all $q^{th}$ orders for the transition:
\begin{equation*}
    d^3 \Sigma : A_{+} \xrightarrow{} \mathcal{R}^q.
\end{equation*}The Wald approach towards black hole entropy is used to calculate the second-order projection of a Gaussian to the third, with no corrections in the metric. Considering the entropy of the lattice as $S_l$, and that of the transformed region as $S_t$, the induced Gaussian hyper-surface implies a third-order correction $d^3 \Sigma$ to the total entropy of the system, $S^*$, which is always increasing:
\begin{equation}
    S^* = S_l + \frac{1}{4}\frac{c^3}{G \hbar}d^3 \Sigma,
\end{equation}where $d^3 \Sigma$ with coordinates ($dr, \ d\theta, \ d\phi$) is proportional to the area and length scale characterizing $p$-brane constraints in the balding and spin-down phase. The pressure $p-p_1= p_2 = p_3$ (transversal stress) measured in an inertial frame of $\dot{S}$ adds a term $\Phi(r)$ with the reparametrization of the free parameter $\lambda$ to time-independent coordinates of $d^3 \Sigma$.

The Wald entropy is given by Von-Neumann phase transition from cylindrical radial, circumferential coordinate $r$ to global co-moving coordinate $r_s$
\begin{dmath}
    S^{*}_{\text{wald}} \mid_{r \xrightarrow{}r_s} = \frac{d\Sigma}{4 G} + 64\pi^2(\sigma(x,y))+ \lambda_{\mu} (log(4 G^2 M^2 \Phi^2(r))-2+2\lambda_{\mu}),
\end{dmath}where Calmet and Kuipers suggested corrections on $\Phi(r)$ from the pressure exerted by non-local quantum effects. With no corrections in ${\mathcal{D}}(\Sigma^3)$, $dM^{(n < 5)}$ is continuous:
\begin{dmath}
    TdS_{l} \equiv  1+ [\lambda_{(1)} \frac{16 \pi}{G M^2} dM^{(1)}]_{r<<r_s} - [\lambda_{(2)} dt^{(0)} \frac{16 \pi}{G M^2} dM^{(2)}]_{r=r_s} +
    (\lambda_{(3)} dt^{(1)} \frac{16 \pi}{G M^2} dM^{(3)}]_{r>>r_s}- [\lambda_{(4)} dt^{(0)} \frac{16 \pi}{G M^2} dM^{(4)}]_{r>>r_s} \dots
\end{dmath}throughout the boundary and tends to infinity for all $t^\mu = [0,1]$ while exhibiting a singular behaviour as \(dM^{(1)} \to 0\), signalling a quantum critical point. This divergence resembles the scaling of thermodynamic quantities near quantum phase transitions, where entropy follows \(S \sim |dM^{(1)}|^{-\nu}\) with critical exponent \(\nu\). Here, \(\nu = 1\) corresponds to mean-field universality, consistent with the linear dependence on \(dM^{(1)}\). This scenario marks the quantum corrections in the vanishing Schwarzschild metric in the mass limit $S_t : dM^{({n\geq 2})} \xrightarrow{}d^3 \Sigma$ and temperature due to Hawking and Unruh-Starobinskii radiation in a spherical lattice of the co-moving radial coordinate $r_s$:
\begin{equation}
     TdS_{t}\ - \ p_{2} \ d^3\Sigma \simeq \ \mid \lambda_{(1)} \frac{16 \pi}{G M^2} dM^{(1)} \mid_{r << r_s}
\end{equation}where $\lambda_{(1)}$ is continuous for orders $q\leq 4$ in the lattice and negligible (quantum gravitational coupling from AdS/CFT correspondence, \(\lambda_{(1)} = \hbar c^5/G^2\) ) for all even orders unless the quintic term in $dM$ breaks the symmetry. Here, \(dM^{(1)} \to 0\) defines the critical temperature (\(T_c\)) for the black hole phase transition.

For a Gaussian $d^3 \Sigma$, the partition function only maps the third-order derivative in volume:
\begin{equation}
    d^{3}\Sigma = \frac{1}{Z} 32 \pi G M^2 \Phi(r) =  \text{constant},
\end{equation}where the membrane vanishes at the extremal boundary $H_{ij}K^{ij}=0$ and $dM=0$. Else, for cases where $dM^{(1)}>0$, instabilities (Aretakis) arise due to the system's inherent symmetry under translations where the constant term indicates fixed entropy density at criticality \( T \to T_c \).

While considering the quantum pressure corrections in the global FLRW scenarios, the energy density of a pressure-less fluid from Friedman's equations could be written as:
\begin{equation}
    \rho(r) = \frac{1}{16 \pi G} \left(\frac{1}{2} \Phi^2(r) + V(\Phi)\right),
\end{equation}where $V(\Phi)$ is the potential energy and $G$ is the gravitational constant. Substituting equation (36) in (35) gives:
\begin{equation}
    TdS_t \equiv 16 \pi G \left(p_2 \frac{2 \Phi(r)}{Z}+\frac{\lambda_{(1)} dM^{(1)}}{(GM)^2}\right),
\end{equation}
where $Z=\Phi^{*}(r) \hat{Z}$ is a conformal mapping interpreted as a third-order correction of $p$-brane constraints: horizon volume to the Wald entropy of horizon area. For example, if the conformal mapping reduces to \( \hat{Z} = \sqrt{-g} \), it recovers the semiclassical entropy. This indicates a universal scaling near criticality, where \(\Phi(r)\) functions as a scaling field, analogous to the magnetic field in the Ising model \cite{mangazeev2010scaling}, while \(dM^{(1)}\) acts as a tuning parameter.

Also, by comparing equations (34) and (35), one could deduce the Aretakis scalar potential for the first-order boundary for the balding phase:
\begin{equation}
    \mid d{\bf u} \mid_{r<<r_s} = -\frac{1}{Z}(\Phi^* (r) +2 V(\Phi) \Phi(r))+V(\Phi),
\end{equation}which is equal to the internal energy $d{\bf u}=TdS_t$ of the brane if $d^3 \Sigma$ is a Gaussian.

However, equation (39) contradicts the extremal boundary data $dM=0$ since the balding phase implies a shearless fluid $\sigma(x,y)=0$ stabilized at a finite horizon boundary $r=r_s$ in equation (35). This means either $\Phi(r)=0$, a continuum from the spin-down phase, or $d{\bf u} \leq 0$ in the balding phase violates conformal symmetry for a quasi-static system, i.e., the entropy (in terms of work done) of the radiating system decreases with the expansion of the membrane at the extremal horizon. So, it is a valid a priori from the spin-down phase that the membrane has unstable curvature data when $\Phi(r)=constant$ and null data when $\Phi(r) < 0$. Meanwhile, the total Wald entropy of the embedding black hole increases in the FLRW system.

\section{Discussion}

This paper investigates black holes' behaviour under extreme conditions, such as near-collapse or phase transitions. The study utilizes three approximations to generate stable quasi-boundary data and simplifies the analysis by considering Schwarzschild black holes in their balding phase. The main results are as follows:

One significant result of this study involves the application of conformal invariance analogy to condensed matter systems exhibiting a continuum of phase transitions. Similar to the phase transition in the Landau-Ginzburg model \cite{tuszynski1984}\cite{doelman1996}, with a quintic term in equation (34), it involves physical data $[K, H, \Sigma]$ and an effective action constrained by surface parameters $\lambda$ and $\Lambda$. This follows that the \(\lambda_{(n)}\)-dependent corrections govern critical exponents (\(\nu, z\)). For instance, the entropy divergence \(dS \sim |dM^{(1)}|^{-\nu}\) suggests \(\nu = 1\), while the dynamic exponent \(z = 2\) arises from the conformal symmetry of the semi-classical metric in equation (4). These exponents characterize universality classes for black hole quantum criticality, analogous to condensed matter systems.

We use the term \say{surface} to refer to the local isometry between two Riemannian manifolds, for example, $X$ and $Y$. Isometry in this context represents a local diffeomorphism map that preserves the boundary data using a parameter when the surface is Riemannian. The advantage of using surface parameters is that their scaling limit at a critical point should exhibit conformal invariance, which locally preserves translation and rotation symmetry. This is shown with a correction in mass limit and temperature using equations (34) and (35) due to Hawking radiation. However, in the mass limit where Aretakis instabilities arise, the associated topology becomes non-local, resulting in the loss of translation and rotational symmetry and a vanishing Schwarzschild metric. This instability, referred to as the \say{soft-hair conjecture with instabilities}, challenges the idea that conformal invariance should apply universally to any generic space-time.

To address this, one could either explore the existence of a smooth map between two topological spaces $\Sigma:X \rightarrow Y$ or the p-brane constraint for a conformally invariant boundary satisfying quasi-equilibrium data for all higher-order corrections. This is, so far, a manipulation of the boundary data as continuous boundaries similar to \cite{miransky1997, miransky2000}. The problem here is primarily on the linear isometries of $N$ lapse functions in the (3+1) lattice model, which should be equivalent to deducing conformal invariance from a cylindrical symmetry.

Conversely, two-dimensional lattice models highlight consistent connections observed in conformal systems undergoing phase transitions \cite{duminil2012}. However, applying this phase symmetry as a generalization to (3+1)-D lattice models implies upholding translational, rotational, and scale symmetry in the boundary data. It is clear that any such consideration, particularly in the extremal mass limit ($dM^{(n<5)}=0$), with an inseparable horizon involving a scalar and a generic spacetime, can lead to instabilities.

While certain mapping also indicates the possibility of stable solutions, as shown in equations (19) and (27), it is essential to acknowledge the limitations and accurately represent the complete behaviours of the horizon. Factors such as specific potential functions, parameter choices, and the formulation of the action may affect the accuracy of the actual representation. Further research and exploration are necessary to address these limitations and provide a more comprehensive understanding of black hole behaviour under extreme conditions.

\section{Acknowledgments}
The authors would like to express their gratitude to Christ University, Bangalore, for funding this research under the Seed Money Project.

\small
\bibliography{ref}

\end{document}